\documentstyle[prb,aps,preprint]{revtex}
\author{Michael Schulz and Steffen Trimper}
\address{Fachbereich Physik\\
Martin-Luther-Universit\"at\\Halle D-06099, Germany}
\title{Collective Diffusion and a Random Energy Landscape}
\begin{document}
\draft
\date{\today}
\maketitle
\begin{abstract}
Starting from a master equation in a quantum Hamiltonian form 
and a coupling to a heat bath we derive an evolution equation for a 
collective hopping process under the influence of a stochastic energy 
landscape. There results different equations in case of an arbitrary 
occupation number per lattice site or in a system under exclusion. Based on 
scaling arguments it will be demonstrated that both systems belong below the 
critical dimension $d_c$ to the same universality class leading to anomalous 
diffusion in the long time limit. The dynamical exponent $z$ can be calculated 
by an $\epsilon = d_c-d$ expansion. Above the critical dimension we discuss 
the differences in the diffusion constant for sufficient high temperatures. 
For a random potential we find a higher mobility for systems with exclusion.\\[1cm]

\end{abstract} 

\section{Introduction}
\noindent Many systems behave on the phenomenological level essentially 
randomly and therefore other approaches for the theoretical treatment 
have to be employed. The randomness, resulting from stochastic 
forces or be intrinsic in the underlying microscopic theory, inevitably 
leads to the description of such systems in terms of probabilities and 
expectation values \cite{vK,sp}. The time development of probability is 
usually found using a master equation. The past years have seen an exciting 
new development based on the observation \cite{doi} of the close relationship 
between the Markov generator of the master equation and a time evolution 
operator acting on a many-particle Fock space \cite{gra,satr}, for some recent 
reviews compare \cite{sti,mg}. The new insight has led to a series of 
remarkable exact solutions for the stochastic dynamics of interacting particle 
systems, for a recent overview see \cite{gs}. Despite of exact results the 
mentioned method has been also fruitful in an approximative description of 
other models such as the facilitated kinetic Ising system as a candidate for 
glassy systems \cite{sctr,sctr1,sctr2} or in branching and annihilation 
random walks \cite{ct}. Whereas the original paper \cite{doi}, 
see also \cite{pe}, are concerned with a mapping of the master equation to a 
representation in terms of second-quantized bosonic operators a great progress 
for exact solvable models had been achieved by mapping to spin-$1/2$ 
Pauli-operators \cite{gs}. This mapping to spin systems applies to processes 
where each lattice site can be occupied by only a finite number of particles. 
Physically, this restriction may be hard-core constraints or fast on-site 
annihilation processes. Obviously, such a mapping simulates the exclusion 
principle for classical lattice models with in a cellular automata.\\
In the present paper the Fock space description is applied for systems far 
from equilibrium which are coupled to a heat bath. In particular, we 
discuss the collective hopping process of a classical many body system,  
coupled to the mentioned heat bath, and under the influence of a random energy 
landscape realized by a stochastic activation energy. The particles making 
random walks have to overcome spatially distributed energy barriers. As the 
consequence the hopping process is accomplished by a competing force field 
which can give rise to anomalous diffusion. Further, the analysis should be 
different considering both cases, the bosonic and the exclusive ones. 
In the first case the particles should find more rapidly the local energy 
minima however because of that their mobility could be reduced. As the 
consequence of the random walk where the particles have to overcome spatially 
distributed energy barriers, the resulting effective force field can give rise 
to anomalous diffusion. It is well known that one of the reasons for an 
anomalous diffusive behavior can be traced back to 
the influence of a stochastic force field below a critical dimension 
\cite{ff,kl}. An alternative way of self induced anomalous diffusion 
had been discussed recently \cite{ss} introducing a feedback coupling between the 
diffusive particle and its local environment. Both, the disorder and the 
memory controlled feedback may lead to a subdiffusive behavior or to 
localization. The analytical approach \cite{ss} could be confirmed by
simulations in one and two dimensions where at the critical dimension $d_c=2$ 
logarithmic corrections in the mean square displacement had been found 
\cite{sst,sst1}.\\
Here, we demonstrate that the Fock space approach leads in both cases, bosonic 
and under exclusion, may lead to anomalous diffusion. Within the long time 
limit and on a large spatial scale both systems belong to the same universality 
class.

\section{Quantum Approach to Nonequilibrium}

\noindent The analysis is based on a master equation 
\begin{equation}
\partial_tP(\vec n,t)=L^{\prime}P(\vec n,t) 
\label{ma}
\end{equation}
where $P(\vec n, t)$ is the probability that a certain configuration 
characterized by a state vector $\vec n = (n_1, n_2 \dots n_N) $ is realized 
at time $t$. There are two special cases, eithe each lattice site is 
occupied by an arbitrary number of particles $n_i = 0, 1, 2 \dots$ or 
as in a lattice gas $n_i = 0, 1$. Further, the occupation numbers $n_i$ 
are considered as the eigenvalues of the particle number operator defined by 
creation operators $d^{\dagger}_i$ or by annihilation operators $d_i$. The 
problem is to formulate the dynamics in such a way that the 
possible realizations for the occupation numbers are taken into account 
explicitly. The situation in mind can be analyzed in a seemingly compact 
form using the master equation in a quantum Hamilton formalism 
\cite{doi,gra,sp,satr,aldr}, for a recent reviews see \cite{mg,gs}. The 
dynamics is determined completely by the form of the evolution operator 
$L^{\prime }$, specified below, and the commutation relations of the 
underlying operators $d^{\dagger}_i$ and $d_i$. Within that approach 
\cite{doi} the probability distribution $P(\vec n,t)$ is related to a state 
vector $\mid F(t) \rangle$ in a Fock-space according to 
$P(\vec n,t) = \langle \vec n\mid F(t)\rangle$. 
The basic vectors $\mid \vec n \rangle$ are composed of the 
operators $d^{\dagger}_i$ and $d_i$. Using the relation 
\begin{equation}
\mid F(t) \rangle = \sum_{n_i} P(\vec n,t) \mid \vec n \rangle
\label{fo2}
\end{equation}  
the master eq. (\ref{ma}) can be transformed into an equivalent 
one in a Fock-space
\begin{equation}
\partial_t \mid F(t)\rangle = L \mid F(t) \rangle
\label{fo1}
\end{equation}
where the operator $L^{\prime}$ in (\ref{ma}) is mapped onto the operator 
$L = \sum \mid \vec m \rangle L^{\prime}_{mn} \langle \vec n \mid $ 
in eq.(\ref{fo1}). It should be emphasized that the procedure is up to now 
independent on the realization of the basic vectors. Originally, the 
method had been applied for the Bose case \cite{doi,gra,pe}. Recently, 
an extension to restricted occupation numbers (two discrete orientations) 
was proposed \cite{sp,satr,aldr}. Further extensions to  
p--fold occupation numbers \cite{sctr2} as well as to models with kinetic 
constraints \cite{sctr3} and to systems with two heat bathes \cite{ssa}  
are possible.\\
As shown by Doi \cite{doi} the average of an arbitrary physical 
quantity $B(\vec n)$ can be calculated by the average of the corresponding 
operator $B(t)$
\begin{equation}
\langle B(t) \rangle = \sum_{n_i} P(\vec n,t) B(\vec n) = 
\langle s \mid B \mid F(t) \rangle 
\label{fo3}
\end{equation} 
with the state function $\langle s \mid = \sum \langle \vec n \mid$. The 
evolution equation for an operator 
$B(t)$ reads now  
\begin{equation}
\partial_t \langle B \rangle = \langle s \mid [B(t),L] \mid F(t) \rangle
\label{kin}
\end{equation}
As the result of the procedure, all the dynamical equations governed by the 
classical problem are determined by the structure of the evolution operator 
$L$ and the commutation rules of the operators. 

\section{Coupling to a Heat Bath}
\noindent The evolution operator for a collective hopping process is different 
for an arbitrary occupation number, denoted as Bose case, or 
an restricted occupation number, denoted as Fermi case. For the last system 
the operator $L_f$ reads \cite{sctr} 
\begin{equation}
L_f = \mu \sum_{i,j} \left( d^{\dagger}_i d_j - (1 - n_i)n_j \right)
\label{fli1}
\end{equation}
where $\mu$ is the hopping rate between adjacent sites $i$ and $j$. 
The occupation number operator $n_i = d_i^{\dagger} d_i$ is related to the 
spin operator by the relation $S_i = 1 - 2n_i$ and the commutation rule 
is $[d_i, d_j] = \delta_{ij} (1 - 2n_i)$.\\
For the Bose case we get
\begin{equation}
L_b = \mu \sum_{i,j} \left( d^{\dagger}_i d_j - n_j \right)
\label{fli1a}
\end{equation}
where $d^{\dagger}_i$ and $d_i$ fulfills the Bose commutation rules. 
A generalization to processes under the coupling to a heat bath 
with a fixed temperature $T$ is discussed in \cite{sctr}. As demonstrated in 
\cite{sctr,st1} the evolution operator has to be replaced by 
\begin{equation}
L_f = \mu \sum_{i,j} \left[ (1 - d_i d^{\dagger}_j)  \exp(-\beta H/2) d^{\dagger}_i d_j \exp(\beta H/2) \right] 
\label{fli4}
\end{equation}
where the hopping rate $\mu$ defines a microscopic time scale; 
$\beta = T^{-1}$ is the inverse temperature of the heat bath and $H$ is the 
Hamiltonian as a measure for the energy. A further generalization is realized 
by introducing different local heat bathes is discussed in \cite{ssa}. 
In the bosonic case the generalization to finite temperatures leads to 
\begin{equation}
L_b = \mu \sum_{i,j} \left[ (1 - \delta_{ij}) \exp(-\beta H/2) d^{\dagger}_i d_j \exp(\beta H/2) \right] 
\label{fli4a}
\end{equation}
Here we study the case that the Hamiltonian $H$ in 
eqs.(\ref{fli4},\ref{fli4a}) is simply given by a stochastic energy 
landscape defined by the energy functional 
\begin{equation}
H = \sum_i \varepsilon_i n_i
\label{sto} 
\end{equation}
Whenever the energy is positive the empty site is energetically favored. 
Further, 
$\varepsilon$ is assumed to be a stochastic local energy the distribution 
of which will be introduced below based on the continuous representation. In 
this manner, the model describes a collective hopping process where the 
jumping particles are subjected to a local random energy $\varepsilon_i$ which 
supports or prevents the hopping process 
with a probability proportional to $\exp(\pm \varepsilon_i/2T)$.  
Taking into account the commutation rules we get in both cases  
\begin{equation}
e^{-\beta H/2} d_i e^{\beta H/2} = d_i e^{\varepsilon_i/2T} \quad\quad
e^{-\beta H/2'} d^{\dagger}_i e^{\beta H/2} = d^{\dagger}_i e^{-\varepsilon_i /2T}
\end{equation}
Using eq.(\ref{kin}) and the algebraic properties of 
Pauli--operators, the evolution equation for the averaged density reads
\begin{eqnarray}
\mu^{-1} \partial_t \langle n_r \rangle &=& \sum_{j(r)} 
[\exp\left((\varepsilon_j-\varepsilon_r)/2T\right) \langle n_j \rangle 
- \exp\left((\varepsilon_r-\varepsilon_j)/2T \right) \langle n_r \rangle \nonumber\\
&-& 2 \sinh (\frac{\varepsilon_j -  \varepsilon_r}{2T}) \langle n_r n_j \rangle      
\label{e1}
\end{eqnarray}
In the Bose case the evolution equation is much simpler.
\begin{equation}
\mu^{-1} \partial_t \langle n_r \rangle = \sum_{j(r)} 
\left[\exp((\varepsilon_j-\varepsilon_r)/2T ) \langle n_j \rangle 
- \exp((\varepsilon_r-\varepsilon_j)/2T ) \langle n_r \rangle \right]
\label{e1a}
\end{equation}
Both equations reflect the conservation of the particle number which will be 
more transparent in a continuum representation. In the special case of a 
constant energy $\varepsilon_r = \varepsilon_j$ it 
results the conventional diffusion equation in a discrete version. When 
the energy changes from site to site the nonlinear eq.(\ref{e1}) is the first 
step of a whole hierarchy of evolution equations. Assuming now smoothly 
changing energy $\varepsilon_r$ and density $n_r$ a gradient expansion is 
appropriate up to the order $l^2$ where l is the lattice size. To make the 
expansion invariant under the underlying rotational symmetry we have to use the 
following identity
\begin{eqnarray}
\sum_{j(r)} \exp((\varepsilon_j-\varepsilon_r)/2T) \langle n_j \rangle &=& 
\sum_{j(r)} \langle n_r \rangle \nonumber\\
&+& \exp(-\varepsilon_r/2T) \sum_{j(r)} \left[ \exp(\varepsilon_j/2T) 
\langle n_j \rangle - \exp(\varepsilon_r/2T) \langle n_r \rangle \right]
\label{con}
\end{eqnarray}
Such an expression reads in a continuous representation including terms of the 
order $l^2$ 
$$
z n({\bf r},t) +  \exp(- \varepsilon({\bf r})/2T) 
\nabla^2 \left[(\exp(\varepsilon({\bf r})/2T) n({\bf r},t) \right]  
$$
with the averaged density $\langle n_r \rangle \equiv n({\bf r}, t)$; $z$ 
is the number of nearest neighbors. After decoupling the nonlinear term 
in eq.(\ref{e1}) and performing the continuous limit the density 
$n({\bf r}, t)$ obeys the following nonlinear diffusion-like equation 
\begin{equation}
\mu^{-1} l^{-2} \partial_t n = \nabla^2 n + n(1 - n) \frac{\nabla^2\varepsilon}{T} 
+ (1 - 2n) \nabla n \cdot \nabla \varepsilon /T
\label{con2}
\end{equation}
In a system with exclusion the density couples in a nonlinear manner to the 
stochastic energy field $\varepsilon({\bf r})$. Due to the exchange coupling of the evolution operator 
$L$ in eq.(\ref{fli1}) the resulting equation (\ref{con2}) is a conserving one 
where the current is given by
\begin{equation}
{\bf j}_f = -\nabla n -  n(1 - n) \frac{\nabla\varepsilon}{T}
\label{con3}
\end{equation}
In the Bose case we find after performing the continuous limit the density 
$n({\bf r}, t)$ obeys the following exact equation 
\begin{equation}
\mu^{-1} l^{-2} \partial_t n = \nabla^2 n + \frac{1}{T} \nabla[ n \nabla \varepsilon ]
\label{con2a}
\end{equation}
The conservation law is manifested in the current 
\begin{equation}
{\bf j}_b = -\nabla n - n \frac{\nabla\varepsilon}{T}
\label{con3a}
\end{equation}
The resulting equation is nothing else as the conventional diffusion equation 
under an additional drift term where the Einstein relation is automatically 
fulfilled. Remark that one can derive a similar equation when the system is 
coupled to two heat bathes with different temperatures. In that case one has 
to replace $\varepsilon({\bf r})/ T$ by $\frac{\nu}{T({\bf r})}$ where $\nu$ 
is the chemical potential and $T({\bf r})$ is the local temperature, see also 
\cite{ssa}. In the Bose case eq.(\ref{con2a}) depends on the density in a 
linear manner. It is of Fokker-Planck-type when the density $n({\bf r},t)$ is 
considered as the single probability distribution to find a particle at site 
${\bf r}$ at time $t$. Such an interpretation is always possible because 
we have not taken into account any interactions. Therefore, the particles are 
independent from each other and the concentration field behaves as the 
probability distribution of a single particle of this system. Different to the 
case of an arbitrary occupation the current ${\bf j}_f$ includes a term 
$n(1 - n)$ which is characteristic for systems with exclusion. Due to the 
exclusion principle the systems reveals a kind of correlation which leads 
even in the mean field limit to a nonlinear current. Following the discussion 
for the Bose case eq.(\ref{con2}) can be interpreted as a nonlinear 
Fokker-Planck-equation for a single particle. The nonlinearity reflects a 
feedback of a particle to itself due to the excluded volume effect.\\
It seems to be more appropriate to introduce the force vector 
${\bf f}({\bf r}) = - \nabla \varepsilon({\bf r})$ the evolution equation 
in the Bose case reads now
\begin{eqnarray}
\mu^{-1} l^{-2} \partial n({\bf r},t) &=& \nabla^2 n - \frac{1}{T} {\bf f} 
\cdot \nabla n - \frac{1}{T} \nabla \cdot {\bf f} n
\label{con4}
\end{eqnarray}
In the Fermi case the corresponding equation is
\begin{eqnarray}
\mu^{-1} l^{-2} \partial n({\bf r},t) &=& \nabla^2 n - \frac{1}{T} {\bf f} 
\cdot \nabla n (1 - 2n) - \frac{1}{T} n(1 -n) \nabla \cdot {\bf f}
\label{con4a} 
\end{eqnarray}
When the force field ${\bf f}({\bf r})$ is a stochastic one the 
system offers anomalous diffusive behavior \cite{ff,kl}.\\
\section{Scaling}
\noindent Now let us discuss both equations when the force field is an 
stochastic pure spatial dependent field, the correlator of which is given by  
\begin{equation}
\overline{f_{\alpha}({\bf r}) f_{}({\bf r}^{\prime})} = 
\phi_{\alpha\gamma}({\bf r} - {\bf r}^{\prime}), 
\quad \overline{f_{\alpha}({\bf r})} = 0
\end{equation}
After averaging over the distribution function of the force field the system 
is homogeneous depending only on the difference of the spatial coordinates. 
The most general form of the function $\phi_{\alpha\gamma}$ 
is given in a Fourier representation by
\begin{equation}
\phi_{\alpha\gamma} = A(\vec q)(\delta_{\alpha\gamma} - n_{\alpha} n_{\gamma}) + 
B(\vec q)n_{\alpha} n_{\gamma} \quad \mbox{with} \quad n_{\alpha} = \frac{q_{\alpha}}{q}
\label{1}
\end{equation}
Introducing dimensionless variables $x \to x \Lambda^{-1},\quad t \to 
t \Lambda^{-z}$, where z is the dynamical critical exponent and further    
$n \to n \Lambda^d$ and according to eq.(\ref{1}) for constant $A$ and $B$ 
${\bf f} \to {\bf f} \Lambda^{d/2} $ we find the critical dimensionality 
$d_c =2$. For $d \le 2$ the term proportional to $\nabla({\bf f} n)$ is 
relevant whereas the additional term in case of exclusive motion $\propto 
n {\bf f} \nabla n$ is only relevant for $d < 2/3$. That means for the 
physical dimension $d \ge 1$ both models belong to the same universality class, 
where only $d \le 2$ the disorder is relevant. Physically the result is obvious 
because in the long time limit and for a large spatial scale the Fermi system 
can be considered to consist of blocks with an increasing size. The larger such a 
block the more irrelevant is to distinguish both cases, arbitrary occupation and 
restricted occupation. In case of $d \le 2$ the system reveals anomalous 
diffusive behavior as it had been demonstrated for a similar model not for the 
density $n({\bf r}, t)$ but for the probability to 
$P$ to find a particle at time $t$ at the point ${\bf r}$. Making the same 
calculation we end up with the flow equations for the dimensionless coupling 
parameters $D = \mu l^2,\quad a = \frac{A}{D^2 T^2}K_d,\quad b= \frac{B}{D^2 T^2} K_d$, with 
$K_d (2\pi)^d$: the volume of the d-dimensional unit sphere and 
$\epsilon = 2 - d$ \quad $\xi = \ln(\frac{\Lambda_0}{\Lambda}$ 
\begin{eqnarray}
\frac{\partial D}{\partial {\xi}} &=& 
D \left[ z - 2 + \frac{ a (d-1)}{d} - \frac{b}{d} \right] \nonumber \\
\frac{\partial a}{\partial {\xi}} &=& a \left[\epsilon - a + 
\frac{b (d-1)}{d}\right] \nonumber\\
\frac{\partial b}{\partial {\xi}} &=& b \left[\epsilon - \frac{a}{d} \right]
\label{rg1}
\end{eqnarray}
In the same manner one can derive an equation for the mean square 
displacement 
$R = \Lambda^2 s(D,a,b)$ with $ s = \langle \bf{r}^2 \rangle$ 
The flow equation can be written as
\begin{equation}
2s = \frac{\partial s}{\partial D} \partial_{\xi} D + 
\frac{\partial s}{\partial a} \partial_{\xi} a + \frac{\partial s}{\partial b} 
\partial_{\xi} b
\label{rg2}
\end{equation}
That equation leads to a scaling behavior of the mean square displacement in the 
vicinity of the fixed points of eqs.(\ref{rg1}). 
In order to keep the diffusivity $D$ fixed to its bare value the 
effective dynamical exponent $z(\xi)$ satisfies 
$z(\xi) = 2 + b(\xi)/d +a(\xi) (1-d)/d $. 
When the disorder is irrelevant the fixed points are $a^{\star} = b^{\star} =0$ 
the exponent is $z=2$. For the fixed point 
$a^{\star} = \varepsilon d,\quad b^{\star} =0$ it results $z = 2 - \varepsilon$ 
and for $a^{\star} = b^{\star} = \epsilon d$ we find 
$z = 2 + O(\epsilon^2)$. These values are well known \cite{ff,kl}. At the 
critical dimension $d_c =2$ we proceed on the following manner. The 
observation time $t$ is related to an initial time $t_0$ by 
\begin{equation}
t = t_0 \exp( \int_0^{\xi} z(\xi^{\prime}) d\xi^{\prime})
\label{rg3}
\end{equation}
Using eqs.(\ref{rg1},\ref{rg2} we can fix the scaling parameter $\xi$ according to 
eq.(\ref{rg3} to be 
$$
\xi \simeq \frac{1}{2} \ln(\frac{t}{t_0}) + \frac{1}{2} \ln(1 + 
\frac{a_0}{2}\frac{t}{t_0})
$$
where $a_0$ is initial value for the parameter $a$. 
From eq.(\ref{rg2}) we find the following behavior for the 
mean square displacement
\begin{equation}
\langle {\bf{r}}^2 \rangle = c_1 \frac{t}{t_0} + c_2 \frac{t}{t_0} 
\ln(\frac{t}{t_0} )
\end{equation}
where $c_1$ and $c_2$ are two non-universal constants. As expected the system 
reveals logarithmic corrections at the critical dimension.

\section{Behavior above the critical dimension}
\noindent The thermalized version of the Fock space representation, 
see eqs.(\ref{fli4},\ref{fli4a}), leads in the limit $T \to \infty$ 
to conventional diffusion. In the high temperature limit the particles 
are able to overcome each barrier and as the consequence of the stochastic 
hopping process one finds diffusive behavior in the long time limit 
independently on the underlying statistics. When the temperature is finite 
there appears a competition between two processes resulting in a different 
behavior for both systems. Bose particles can easily find a minimum within the 
energy landscape defined by the stochastic force. Particles with exclusion 
have to search for a longer time and on a larger scale to reach an appropriate 
potential minimum. From here one would conclude to an enhanced diffusivity. On 
the other hand, the mobility of Bosons is eventually reduced because they 
find more rapid a stable minimum. Due to the established universality for 
low dimensions a variation of the behavior should be only observed 
above the critical dimension. In this regime conventional perturbation theory 
should be applicable. Let us therefore present lowest order corrections to 
the the diffusion parameter $D$. The effective diffusivity is defined by
\begin{equation}
D_{eff} = \left | \frac{\overline{\partial n^{-1}(\vec q, \omega)}}{\partial q^2} \right |_{q=0,\omega=0}
\label{sto2}
\end{equation}
As well as the Bose and the Fermi system lead in second order,  
proportional to $\frac{1}{T^2}$, to nontrivial corrections which are also 
manifested in the averaged density $\overline{n({\bf r},t)}$ or the averaged 
correlation function $\overline{n({\bf r},t) n({\bf r}^{\prime},t^{\prime})}$. 
Indeed, the Fermi system offers additional terms for the density or the 
correlation function compared with the Bose case. However those terms does not 
contribute at zero wave vector and hence there are not relevant corrections to 
the divergent part of $D_{eff}$ for $d \le d_c$. Above $d_c$ the behavior of 
the effective diffusion coefficient can be estimated using a perturbative 
approach around the homogeneous solution 
denoted by $\overline{n}$. We get  
\begin{eqnarray}
D_{eff}^f  &=& D_{eff}^b + \frac{(1- \overline{n}) \overline{n} }{D T^2} I 
\nonumber\\
\mbox{with}\quad I &=& \frac{4 K_d}{d} I_1 [B - A (d-1)]  
\label{sto3}
\end{eqnarray}
$I_1$ can be expressed by a momentum integral which is always positive in the 
mesoscopic regime $\Lambda > l$. For $B - A (d-1) > 0$, realized for a 
pure potential field ($B$ is the relevant variable, see eq.(\ref{1})), 
eq.(\ref{sto3}) leads to 
\begin{equation}
D_{eff}^f  > D_{eff}^b 
\label{fl2}
\end{equation}
Remark that the correction to the bare diffusion coefficient $D$ is of the 
order $(1 - 2 \overline{n})^2$, that means for the half-filled case there are 
no corrections. That reasonable result should be also valid in a more refined 
approach.\\ 
Because the homogeneous solution is not necessary a stable one we can also 
estimate the behavior using linear stability analysis around the stationary 
solution denoted as $n_s({\bf r})$. Let us introduce 
$n({\bf r},t) = n_s({\bf r}) + y({\bf r},t)$ then the correction 
$y({\bf r},t)$ fulfills in the Bose case the equation
\begin{equation}
\partial_t y = D \nabla^2 y + \frac{D}{T} \nabla(y \nabla \varepsilon_b) \quad 
\mbox{with}\quad 
{\bf f}({\bf r}) = - \nabla \varepsilon_b({\bf r}) 
\label{fl}
\end{equation}
Here $\varepsilon_b({\bf r}) = \varepsilon({\bf r}) - v$ is the true 
stochastic potential introduced by eq.(\ref{sto}) and 
$v$ plays the role of the chemical potential which regulates the occupation 
number. In case of the exclusion model the deviation from the stationary 
solution $y({\bf r},t)$ satisfies the same equation however one has to 
replace the potential in the Bose case, given in eq.(\ref{fl}), by another 
effective potential
\begin{equation}
\varepsilon_b({\bf r}) \rightarrow \varepsilon_f({\bf r}) = 
2 T \ln \left[ \frac{\cosh \left(\frac{\varepsilon({\bf r}) - v}{2 T}\right)}
{\cosh\frac{v}{2 T}}\right] 
\label{e3}
\end{equation}
We have gauged the potentials so that for $\varepsilon_f({\bf r}) = 0$
also $\varepsilon({\bf r}) = 0$. The hopping particles under exclusion are 
subjected to the modified stochastic energy landscape given by 
$\varepsilon_f$. Expanding $\varepsilon_f$ in terms 
of $\varepsilon$ we find the relation
\begin{equation}
\varepsilon_f({\bf r}) \simeq - \tanh(\frac{v}{2T}) \varepsilon({\bf r})
\label{e2}
\end{equation}
From here it results   
\begin{equation}
\overline{\varepsilon_f({\bf r})\varepsilon_f(0)} \simeq 
\tanh^2 (\frac{v}{2T})\quad\overline{\varepsilon_b({\bf r})\varepsilon_b(0)}
\label{fl1}
\end{equation}
The effective correlator of the disorder in the Fermi system is drastically 
decreased in comparison to the Bose case. This result is compatible 
with the previous discussion leading to eqs.(\ref{sto3},\ref{fl2}). In 
particular in the vicinity of half-filling (where the chemical potential $v$ 
is zero) the influence of the disorder is very weak. This special case 
corresponds to vanishing linear term expanding eq.(\ref{e3}) 
according to powers of $\varepsilon$. In the leading order we obtain
$$  
\varepsilon_f({\bf r}) \simeq \frac{\varepsilon^2({\bf r})}{4T} 
$$
Different to the Bose case the effective stochastic potential $\varepsilon_f$, 
eq.(\ref{e3}), is always positive definite, that means all the deep negative 
minima of the original stochastic potential become maxima and therefore they 
are not more available in case of Fermi system. Obviously, they 
are already occupied and hence they are not accessible for particles. 

\section{Conclusions}
 
\noindent In the present paper the collective hopping process on 
a lattice is studied systematically when the particles are subjected to a  
random energetic landscape manifested by a stochastic energy profile. In 
particular, we have taken into account both cases, each lattice site is only 
occupied by one particle or each site can absorb an arbitrary number of 
particles. Physically, one expects different behavior. Whereas in the situation 
under exclusion a particle should spend more time for searching an appropriate 
energy minimum within the stochastic energy the bosons tend to reduce their
mobility because they remain for a longer time in the local minima. 
A further influence on the motion of the particles is given by the coupling 
to a heat bath which supports the tendency that the system equilibrates.
Starting on a master equation in a second quantized form both cases can be 
easily realized in terms of Bose-operators or spin-$1/2$ Pauli-operators. 
The annihilation and creation process of particles leads in both cases 
to a density gradient characteristic for a random walk. Due to the additional 
coupling to a stochastic energy each particle can not follow that gradient simply 
but it has to overcome an energy barrier at its starting point and at its 
end point. There appears a conflicting situation that a 
particle follows the density gradient but the energy at the starting point 
is higher than at the end point. In this manner it will jump from an occupied 
to an empty site however under mobilizing a higher amount of energy 
(lower temperature). The other situation consists of the fact that a particle 
follows the density gradient and the energy 
barrier at the starting point is lower than at the end point (high temperature 
regime). In this case the hopping process is highly supported by the energy 
landscape whereas in the previous one the process is restricted. As the 
consequence anomalous diffusive behavior should be realized below the critical 
dimension.\\
In the paper we have demonstrated that the Bose-as well as the Fermi-system 
belong below the critical dimension to the same universality class within the 
long time limit and on a large spatial scale. For an increasing scale the 
system can be considered consisting of blocks with an increasing number of 
particles. Thus, the cases 
of restricted and unrestricted occupation number per lattice site should be 
irrelevant. Despite of the universality the density and the correlation 
function of both systems are different, in particular for an intermediate 
interval. In particular, we have discussed the situation above the critical 
dimension where the diffusion constant can offer different behavior in both 
cases.

\end{document}